%% file: main.tex
\documentclass[aps, prd, twocolumn, amsmath, floats,floatfix, superscriptaddress, nofootinbib]{revtex4}
 \usepackage{natbib}
\usepackage{graphicx,color}
\usepackage{amsmath,amssymb}
\usepackage{verbatim}
\usepackage{float}
\usepackage{wasysym}
\usepackage{amssymb,graphicx,subfigure}
\usepackage{epsfig}
\usepackage{psfrag}
\usepackage{dsfont}
\usepackage{amsfonts}
\usepackage{mathrsfs}
\usepackage{multirow}
\usepackage{times}
\usepackage{bm}
\usepackage{hyperref}
\usepackage{xspace}
\usepackage[export]{adjustbox}
\usepackage{orcidlink}
\usepackage{soul}
\hypersetup{
  colorlinks=true,        
  linkcolor=blue,         
  citecolor=cyan,         
  }
\usepackage{pifont}
\usepackage[normalem]{ulem}                                                                                                                                                         
\graphicspath{{./}}
\input kigou.tex

\begin{document}
\title{Gravitational-wave imprints of non-convex dynamics in binary neutron star mergers}
\author{Giuseppe Rivieccio \orcidlink{0009-0009-9456-6382}}
\affiliation{Departament d'Astronomia i Astrofísica, Universitat de València, C/ Dr Moliner 50, 46100, Burjassot (València), Spain}
\author{Davide Guerra \orcidlink{0000-0003-0029-5390}}
\affiliation{Departament d'Astronomia i Astrofísica, Universitat de València, C/ Dr Moliner 50, 46100, Burjassot (València), Spain}
\author{Milton Ruiz
\orcidlink{0000-0002-7532-4144}}
\affiliation{Departament d'Astronomia i Astrofísica, Universitat de València, C/ Dr Moliner 50, 46100, Burjassot (València), Spain} 
\author{José A. Font
\orcidlink{0000-0001-6650-2634}}
\affiliation{Departament d'Astronomia i Astrofísica, Universitat de València, C/ Dr Moliner 50, 46100, Burjassot (València), Spain} 
\affiliation{Observatori Astronòmic, Universitat de València, C/ Catedrático José Beltrán 2, 46980, Paterna (València), Spain}
\date{today}

\begin{abstract}

Explaining gravitational-wave (GW) observations of binary neutron star (BNS) mergers requires an understanding of matter beyond nuclear saturation density. Our current knowledge of the properties of high-density matter relies on electromagnetic and GW observations, nuclear physics experiments, and general relativistic numerical simulations. In this paper we perform numerical-relativity simulations of BNS mergers subject to non-convex dynamics, allowing for the appearance of expansive shock waves and compressive rarefactions. Using a phenomenological non-convex equation of state we identify observable imprints on the GW spectra of the remnant. In particular, we find that non-convexity induces a significant shift  in the quasi-universal relation between the peak frequency of the dominant mode and the tidal deformability (of order $\Delta f_{\rm peak}\gtrsim 380\,\rm Hz$) with respect to that of binaries with convex (regular) dynamics. Similar shifts have been reported in the literature, attributed however to first-order phase transitions from nuclear/hadronic matter to deconfined quark matter. We argue that the ultimate origin of the frequency shifts is to be found in the presence of anomalous, non-convex dynamics in the binary remnant. 

 
\end{abstract}

\maketitle

\section{Introduction}\label{sec:intro} 
Gravitational waves (GWs) from binary neutron star (BNS) mergers encode key information about the nature of matter above nuclear saturation density~($n_0= 0.15\pm 0.01\,\rm fm^{-3}$). Following merger, the bulk of the GW energy is emitted and reaches values of $\sim 0.1M_\odot c^2$~\cite{Zappa:2017xba}. This energy is emitted at frequencies $\gtrsim 2\,\rm kHz$ and would be observable with the third-generation detectors Einstein Telescope and Cosmic Explorer~\cite{Maggiore:2019uih,Branchesi:2023mws,2021arXiv210909882E}. The GW spectrum is directly linked to  properties of NS~\cite{Bauswein:2019ybt} and can be used to impose tight constraints on the equation of state (EoS), complementary to those from electromagnetic (EM) observations and  heavy-ion experiments~\cite{Glendenning:1997wn,Heiselberg:1999mq,Haensel:2007yy}.  These constraints can in turn be used to infer a number of key properties of a NS, as e.g.~the mass-radius relationship, the tidal deformability, or the moment of inertia. In particular, numerical work has shown that the GW spectra of the remnant is characterized
by the presence of distinctive peaks associated with different oscillation
modes (see e.g.~\cite{Soultanis:2022,Topolski:2023} and references therein). The frequencies of various such modes, e.g.~the peak frequency at merger $f_{\rm peak}$ and the quadrupolar mode frequency $f_2$, have been found to be related quasi-universally with the tidal deformability parameter $\Lambda$ characterising the quadrupolar deformability of an isolated NS. 

Numerical relativity simulations of BNS mergers have also revealed that following merger the temperature inside the densest parts of the binary remnant remains below $T\lesssim 10\,\rm MeV$, while hot patches of matter with $T\gtrsim 50\,\rm MeV$  eventually appear, triggering the formation of a hot annulus~(see~e.g.~\cite{Kastaun:2016yaf,Hanauske:2016gia,Raithel:2023zml,DePietri:2019mti}).  Through angular momentum transport, matter in the outer region of the system gains enough rotational energy to be ejected, while the inner part contracts to form a central core, which may undergo a transition to quark-gluon plasma or other exotic states~\cite{Orsaria:2019ftf,Most:2018eaw,Chamel:2012ea}. It has also been shown~(see e.g.~\cite{Kochankovski_2022, Raduta_2021, Raduta_2022}) that particle production such as hyperons,  a process that becomes relevant as the temperature increases, can trigger a substantial drop in the thermal pressure of the binary remnant. 

Although there have been considerable efforts into the theoretical understanding of the EoS beyond nuclear saturation density, systematic calculations of matter properties at densities larger than $n_0$ based on quantum chromodynamics (QCD) are still not possible. Therefore, many properties of phase transitions remain unclear, e.g. the threshold temperature or the  densities at which the system undergoes a phase transition are unknown~\cite{Blacker:2020}. So far, QCD calculations assuming a vanishing baryonic chemical potential $\mu$ predict that  
{a smooth crossover transition} will take place at a temperature of $T = 154\pm 9\,\rm MeV$~\cite{Bazavov:2012vg,Borsanyi:2013bia,HotQCD:2014kol}.  Unfortunately, at $\mu\neq 0$ only perturbative calculation or phenomenological QCD models exist~\cite{Kruger:2013kua,Kurkela:2014vha,Lovato:2022vgq}, and  they are not, in particular, applicable in regions of first-order phase transitions. 

The possible appearance of phase transitions from nuclear hadronic matter into quark-gluon plasma or into matter phases containing exotic particles (e.g.~hyperons) in BNS mergers may modify the stability, dynamics and final fate of the remnant, and thus the associated GW signal. Numerical studies have sought to identify the imprints of the first-order hadron-quark phase transition on the GWs~(see e.g.~\cite{Radice:2017,Most:2018eaw,Bauswein:2018bma,Blacker:2020,Liebling:2020dhf,Ujevic:2022nkr}). In particular, the BNS merger simulations of~\cite{Bauswein:2018bma, Blacker:2020} showed that if the remnant undergoes a {\it strong} first-order phase transition to deconfined quarks, the  dominant GW frequency at merger $f_{\rm peak}$ exhibits a significant deviation from a quasi-universal (i.e.~EoS-insensitive) relation with the tidal deformability $\Lambda$, an effect that could be observationally identified. We note that phase transitions soften\footnote{{In a NS modeled with a stiff (soft) EoS the pressure increases promptly with density. Hence, its core is relatively resistant (susceptible) to compression giving rise to stars with larger (smaller) radii.}} the EoS at merger, which potentially can modify the ejecta properties and, hence, any EM counterpart~\cite{Radice:2017,Most:2018eaw,Bauswein:2018bma,Fujibayashi:2017xsz}.

Physical processes involving  high-density matter where the system undergoes a phase transition to exotic states also affect the monotonic increase of the speed of sound with density. In particular, at densities above $n_0$ monotonicity is lost~\cite{Bazavov:2012vg,Borsanyi:2013bia,HotQCD:2014kol,Bedaque:2015,McLerran_2019,Tan_2022,Somasundaram_2023,Yao:2023,Mroczek:2023,Dey:2024}. The non-monotonicity of the sound speed can
also result from the behaviour of the
adiabatic index at such densities~\cite{Haensel:2004}.
The speed of sound is closely related to the so-called {\it convexity} of  the EoS. Namely, the convexity of a thermodynamical system is determined by the sign of the so-called {\it fundamental derivative}~$\mathcal{G}$ on the $p-V$ plane~\cite{Menikoff:1989ka,Toro09}, a quantity directly connected to the derivative of the speed of sound. Here $p$ and $V=\rho_0^{-1}$ are the pressure and the specific volume, respectively, with $\rho_0$ the rest-mass density. When $\mathcal{G}> 0$ isentropes on the $p-V$ plane are convex and  the dynamics of the system involves compressive shocks and expansive rarefaction waves~\cite{Thompson:1971}. Such physical systems are said to be convex.  By contrast, when $\mathcal{G}< 0$ the dynamics becomes ``anomalous'', involving expansive shocks and compressive rarefaction waves, and the system is said to be non-convex.
 
In the presence of phase transitions to exotic components, the fundamental derivative can indeed be negative, implying that the EoS should be non-convex in that regime. This would lead to non-convex, atypical dynamics. In particular, at the phase transition the fundamental derivative is discontinuous, i.e.~in the continuous limit there is a single point along an isentrope where $\mathcal{G}< 0$. However, state-of-the-art numerical simulations of BNS mergers employ microphysical, finite-temperature EoS tables constructed using data from observations and nuclear physics~\cite{Typel:2013rza,Oertel:2016bki,stellarcollapse}. Because of the {\it discrete} nature of the tables, it is unlikely that tabulated points coincided with the locus of the discontinuity of the fundamental derivative which may cause non-convex behaviour to spread  spuriously along neighboring points. Besides, finite difference derivatives of the thermodynamical variables may also induce spurious oscillations in $\mathcal{G}$, artificially triggering non-convex dynamics in a finite region. An example of {\it numerical} loss of convexity associated with insufficient thermodynamic discretization of some tabulated EoS when the adiabatic index is non-constant was reported in~\cite{Vaidya:2015}.

Studies of non-convex, relativistic fluid dynamics and magnetohydrodynamics (MHD) have been presented in~\cite{Ibanez:2013,Ibanez:2015,Ibanez:2017xrx,Aloy_2019,Marquina:2019,Berbel:2023qem,Berbel:2023}. Particularly relevant for the topic discussed in this paper are the results reported in~\cite{Ibanez:2017xrx,Aloy_2019,Berbel:2023qem}. In particular~\cite{Aloy_2019} probed the effects of a non-convex EoS on the dynamics of both spherically-symmetric and uniformly-rotating NS undergoing gravitational collapse to black holes (BHs). 
The stars were evolved assuming a phenomenological $\Gamma-$law EoS first proposed in~\cite{Ibanez:2017xrx} for which the adiabatic index $\Gamma$  depends on the rest-mass 
density in a way which leads to  $\mathcal{G}<0$ in some regions of the rest-mass density distribution. The results of~\cite{Aloy_2019} showed that a non-convex EoS has a major effect on the dynamics of gravitational collapse, accelerating the onset of the collapse and leaving distinctive imprints on the GW signal, as compared to the case of a convex dynamics. Moreover, \cite{Aloy_2019}
suggested that to properly capture the transition from nuclear matter to exotic matter states, EoS tables should be more densely populated with nodal points along regions where the fundamental derivative displays large variations, specially when these variations drive negative values of $\mathcal{G}$. We also note that~\cite{Aloy_2019} pointed out that convexity across phase transitions may occasionally be recovered numerically if the singularities in the Gibbs or Helmholtz free energies are removable. Thermodynamic consistency requires the Gibbs free energy to be a jointly concave function. This requirement can be imposed by convolving the Gibbs free energy with a non-negative smoothing function, which smoothes out singularities along phase transitions~\cite{Menikoff:1989ka}. Recently, an analytic model for tabulated EoS that focuses on the modelling of phase transitions through a thermodynamically adaptive piecewise-polytropic approximation has been reported~\cite{Berbel:2023qem}. This method  is able  to reproduce the non-convex behavior of several nuclear EoS.
%
%
\begin{figure}[t]
    \includegraphics[scale=0.56]{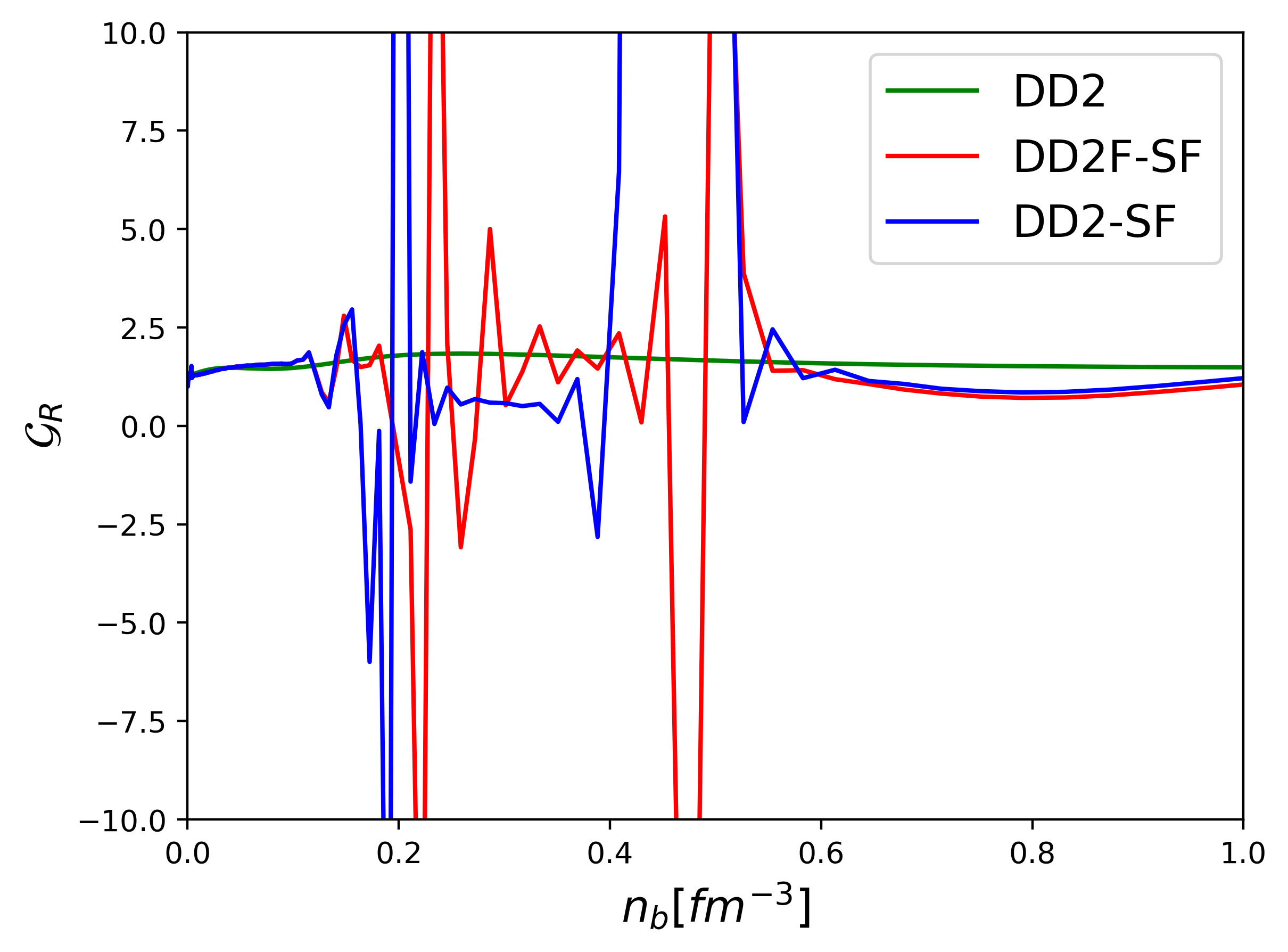}
  \caption{Relativistic fundamental derivative $\mathcal{G}_R$ as a function of baryon density $n_b$ for a few selected EoS with (DD2F-SF and DD2-SF) and without (DD2) a first-order hadron-quark phase transition.} 
\label{plot:phase_tran}
\end{figure}

In this work we analyze the possible repercussion the use of a non-convex EoS may have on the dynamics of BNS mergers, focusing in particular on the evolution of the post-merger remnant. These astrophysical systems offer a perfect framework to study the impact of non-convex thermodynamics on their (hydro) dynamics, as they are characterized by the presence of high-density regions (above nuclear saturation density) where the dynamics may become non-convex.
Moreover, they also allow to investigate the potential influence of non-convex effects on multi-messenger observables such as GW and EM waves. Here, we perform a numerical-relativity survey of BNS mergers in quasi-circular orbit that merge and form a transient remnant. The stars are assumed to be irrotational and are modeled using piecewise-polytropic representations of different microphysical, nuclear EoSs. The effects of non-convex thermodynamics in the BNS mergers are incorporated using the same phenomenological EoS employed in~\cite{Ibanez:2017xrx,Aloy_2019}, i.e.~a $\Gamma-$law EoS allowing for shock heating.  Following~\cite{Ibanez:2017xrx} we assume that  the adiabatic index $\Gamma$ is not constant but depends on the rest-mass density (see below). This allows us to mimic some key  features of tabulated, nuclear-matter EoS such as the non-monotonic dependence of the speed  of sound (or the adiabatic index) with the rest-mass density~\cite{Lim:2019ozm} and, thus, the appearance of non-convex dynamics.
We find that the use of a non-convex EoS does influence the post-merger dynamics in a significant way. In particular, non-convex dynamics can strongly impact the frequency of the peak first visible in the GW spectra right after merger, $f_{\rm peak}$. Depending on the parameters of our EoS, deviations from a $\Lambda-f_{\rm peak}$ quasi-universal relation can be {in magnitude} as large as $\Delta f_{\rm peak}\geq 380\,\rm Hz$ with respect to that of binaries with pure convex evolution. 
Such frequency shifts are reminiscent of the results reported by~\cite{Bauswein:2018bma,Blacker:2020}, where they were attributed as due to a strong first-order phase transitions from nuclear/hadronic matter to deconfined quark matter. We argue that the explanation for the observed frequency shift is to be found in the presence of anomalous, non-convex dynamics in the binary remnant. Our explanation does not exclude the interpretation of the shifts as due to a first-order phase transition, as the dynamics is indeed non-convex there. The explanation based on the existence of a first-order phase transition can be regarded as a particular manifestation of a more general reason, namely the possible non-monotone behaviour of the EoS of NSs above nuclear saturation density.

The rest of the paper is organized as follows:
Section~\ref{sec:Non-convexity_descrip} presents a brief summary of non-convex thermodynamics and of the phenomenological EoS used in this study. The description of the numerical methods employed in the simulations of BNS mergers, including also the initial data and the grid structure, are given in Sections~\ref{sub:initialdata} and 
\ref{sub:methods}. We present our results in Section~\ref{sec:discussion} and summarize our findings and conclusions in Section~\ref{sec:Conclusion}. Finally,  Appendix~\ref{sec:appendix} gathers further evidence from additional simulations to validate our main findings. 

%
%
\section{Non-convex thermodynamics and EoS}
\label{sec:Non-convexity_descrip}
The study of the physical properties of BNS merger remnants requires of the understanding of the EoS at densities typically higher than nuclear saturation density. As discussed above, at such densities the system may develop non-convex dynamics. In the following, we summarize key properties of non-convex dynamics, referring the reader to~\cite{Ibanez:2017xrx,Aloy_2019} for further details.

The  convexity properties of Newtonian hydrodynamical flows is determined by the EoS through the concept of the fundamental derivative~\cite{Bethe1998,zel1946possibility,Thompson:1971} defined as
\begin{eqnarray}
     \mathcal{G}\equiv -\frac{1}{2}\,V\,
     \frac{
   {\left.\frac{\partial^2 p}{\partial V^2}\right|_s} 
   }
   {
  {\left.\frac{\partial p}{\partial V}\right|_s} 
  }
  \,,
     \label{eq:fundD}
\end{eqnarray}
with $s$ being the specific entropy. A change in the sign of the fundamental derivative measures the convexity of the isentropes on the $p-V$ plane.  When $\mathcal{G}>0$  the system is convex and its dynamics involves expansive rarefaction waves and compressive shocks. This is the usual regime in which
many astrophysical scenarios develop. By contrast,  when $\mathcal{G}<0$ the system is non-convex and its (anomalous) dynamics involves  compressive  rarefaction waves and  expansive shocks. This non-standard behavior has been experimentally observed in  transonic and mildly supersonic fluids~\cite{cinnella_congedo_2007,10.1063/1.3657080}. Fluids attaining negative values of the fundamental derivative are called Bethe-Zel’dovich-Thompson fluids, after~\cite{Bethe1998,zel1946possibility,Thompson:1971}.
%
\begin{figure}
    \includegraphics[scale=0.70]{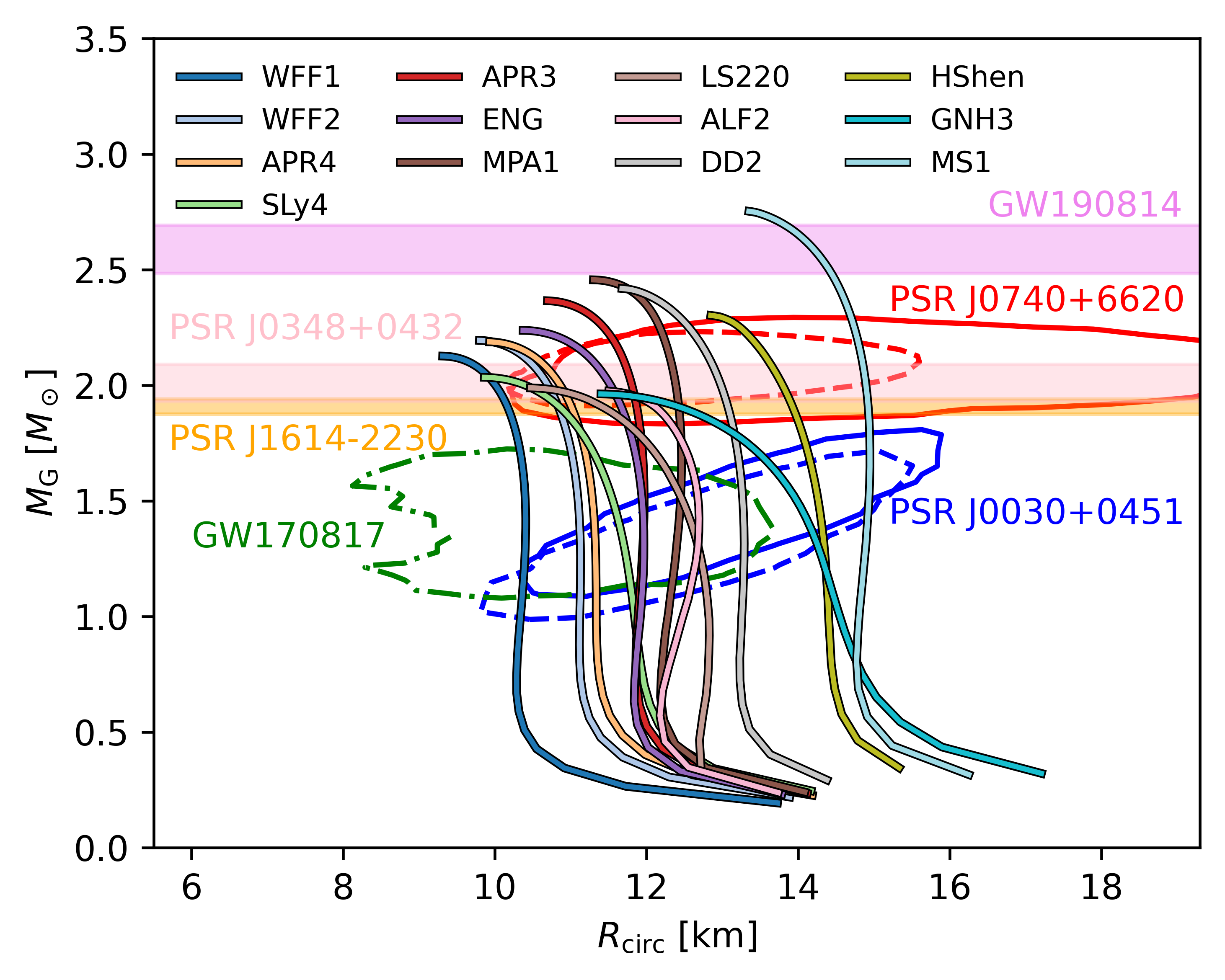}
  \caption{Gravitational mass $M_G$ vs circumferential radius $R_{\rm circ}$
for all the EoSs listed in Table~\ref{table:Iparamenters} along with the observational 
NS mass constraints within 95\% confidence levels from the measurements of pulsars 
from NICER/XMM-Newton~\cite{Fonseca:2016tux,Riley:2021pdl,Riley:2019yda}, and the BNS observations from  the LIGO/Virgo/KAGRA collaboration~\cite{LIGOScientific:2017vwq,LIGOScientific:2020zkf}.} 
\label{plot_MvsR}
\end{figure}

Following~\cite{Menikoff:1989ka,Ibanez:2017xrx}, the speed of sound $c_s$ can be related to the fundamental derivative as
\begin{align}
\mathcal{G}&=1+\left.\frac{\partial\, \ln{c_s}}{\partial\, \ln{\rho_0}}\right|_s \,.
\label{eq:fund_der_w_sound}
\end{align}
Therefore, the fundamental derivative becomes negative when ${\partial\,{\rm ln }c_s}/{\partial\,{\rm ln }\rho_0}|_s <-1$. 
A generalized fundamental derivative for {\it relativistic} fluids was found in~\cite{Ibanez:2013}, given by
\begin{equation}
\mathcal{G}_R=\mathcal{G}-\frac{3}{2}\,c^2_{s(R)}\,,
\label{eq:fund_rel}
\end{equation}
where $c_{s(R)}$ is the relativistic speed of sound that can be related to $c_s$ through $c_s^2=h\,c^2_{s(R)}$ where $h=1+\epsilon+{p}/{\rho_0}$ is the specific enthalpy and $\epsilon$ is the specific internal energy density.

\begin{center}
  \begin{table}
    \caption{Summary of the initial properties of the BNS configurations.
      We list the EoS, the gravitational mass $M[M_\odot]$, the compactness 
      $\mathcal{C}\equiv M/R_{\rm eq}$ and the tidal deformability
      $\Lambda = (2/3)\kappa_2\,\mathcal{C}^{-5}$ for each individual star.
      Here $R_{\rm eq}$ is the equatorial coordinate radius, and $\kappa_2$ is the second Love number. The ADM mass $M_{\rm ADM}[M_\odot]$, the ADM angular momentum
      $J_{\rm ADM}[M_\odot^2]$, the angular velocity $\Omega [\rm krad/s]$, for
      an initial binary coordinate separation of $\sim 44.3\,\rm km$, and the initial maximum value 
      of the rest-mass of the system $\rho_{0,\rm max}[\rm g\,cm^{-3}]$. In all cases
      the NS has a rest-mass $M_0=1.4M_\odot$. 
      \label{table:Iparamenters}}
    \begin{tabular}{cccccccc}
      \hline
      \hline
       EOS      & $M$      & $\mathcal{C}$   &  $\Lambda$  &$M_{\rm ADM}$  & $J_{\rm ADM}$ & $\Omega$ & $\rho_{0,\rm max}$\\
      \hline
       WFF1     &   1.26   & 0.18   & 406.07  & 2.50    & 6.45       & 1.76 & $10^{15.01}$\\
       WFF2     &   1.27   & 0.17   & 1115.06 & 2.52    & 6.54       & 1.76 & $10^{14.90}$\\
       APR4     &   1.28   & 0.17   & 440.75  & 2.52    & 6.56       & 1.77 & $10^{14.93}$\\
       SLy4     &   1.28   & 0.16   & 511.70  & 2.54    & 6.62       & 1.77 & $10^{14.93}$\\
       APR3     &   1.28   & 0.16   & 620.00  & 2.53    & 6.61       & 1.77 & $10^{14.86}$\\
       ENG      &   1.28   & 0.16   & 636.35  & 2.53    & 6.60       & 1.77 & $10^{14.87}$\\
       MPA1     &   1.28   & 0.15   & 784.52  & 2.54    & 6.64       & 1.77 & $10^{14.81}$\\
       LS220    &   1.29   & 0.15   & 899.05  & 2.55    & 6.69       & 1.77 & $10^{14.84}$\\
       ALF2     &   1.29   & 0.15   & 941.42  & 2.54    & 6.66       & 1.77 & $10^{14.79}$\\
       DD2      &   1.29   & 0.13   & 1113.92  & 2.56    & 6.73       & 1.78 & $10^{14.76}$\\
       HShen    &   1.30   & 0.14   & 1633.24 & 2.58    & 6.82       & 1.78 & $10^{14.69}$\\
       GNH3     &   1.30   & 0.13   & 1371.15 & 2.58    & 6.81       & 1.78 & $10^{14.77}$\\
       MS1      &   1.30   & 0.13   & 2020.75 & 2.58    & 6.83       & 1.79 & $10^{14.63}$\\ 
       \hline
    \end{tabular}
  \end{table}
\end{center}

Fig.~\ref{plot:phase_tran} displays the relativistic fundamental derivative for a set of selected EoS from the {\tt CompOSE}~database~\cite{compose} as a function of baryon density $n_b$. EoS DD2F-SF and DD2-SF include a phase transition to hadron-quark matter while EoS DD2 does not. The presence of the phase transition induces the loss of monotonicity of the speed of sound which, subsequently, triggers the loss of convexity.
Indeed, at densities where the EoS suffers a phase transition, the relativistic fundamental derivative $\mathcal{G}_R$ (as well as the Newtonian one) becomes negative. Notice that, as pointed out in~\cite{Aloy_2019}, the oscillating behavior in $\mathcal{G}_R$ is a numerical artifact due to the computation of the fundamental derivative using a discrete EoS table. This causes the non-convex behaviour to spread spuriously along neighboring points.  

Using the above definitions, the effects of a non-convex EoS on the stability and dynamics of isolated NS were probed in~\cite{Aloy_2019} employing a phenomenological Gaussian $\Gamma$-law (GGL) EoS $p = (\Gamma-1)\,\rho_0\,\epsilon$. Here $\Gamma$ is an effective thermal index that, to mimic its  dependency on the nucleon effective mass
for densities above half nuclear saturation~\cite{Lim:2019ozm}, is given by~\cite{Ibanez:2017xrx}
\begin{align}
\Gamma= \Gamma_{\rm th} + \left(\Gamma_1-\Gamma_{\rm th}\right)\,
\rm{exp}\left[-\frac{(\rho_0-\rho_1)^2}
{\Sigma^2}\right]\,,
\label{eq:Gamma_rho}
\end{align}
where $\Gamma_{\rm th}$, $\Gamma_1$, $\rho_1$, and $\Sigma$ are free constant parameters. The results reported by~\cite{Aloy_2019} indicate that a non-convex dynamics can  accelerate the onset of the collapse of the NS to a BH with respect to that of a convex dynamics. Non-convexity also leaves an imprint on the GW signal, amplifying the amplitude of the GWs emitted by the collapsing star. The maximum amplitude is about twice as large as in the convex case.  These imprints are large enough to be detectable by third-generation,  ground-based detectors. 
%
%
\begin{figure}[t]
\includegraphics[scale=0.58]{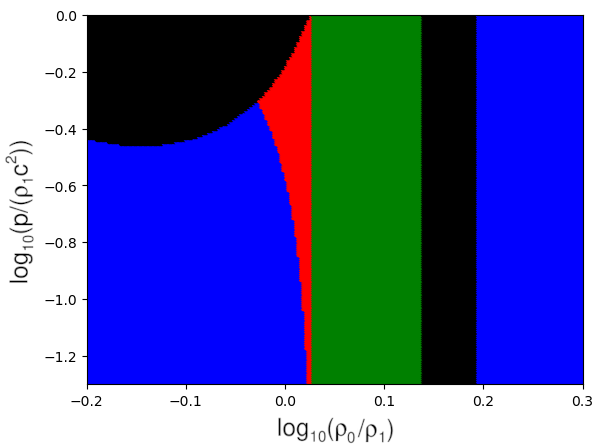}
  \caption{Convexity behavior of the GGL EoS for the canonical values  of the
  free parameters used in our simulations. We set $\Gamma_{\rm th}=1.8$ (as in typical 
  BNS simulations; see e.g.~\cite{Takami:2014tva,DePietri:2019mti,Guerra:2023}), $\Gamma_1=3.0$, $\rho_1=9.1\times 10^{14}\,\rm g\,cm^{-3}$, and $\Sigma=0.35\rho_1$. 
  The blue area corresponds to the region of the parameter space where the EoS is convex 
   (i.e.~$\mathcal{G}>0$ and $\mathcal{G}_R>0$).  The green region  corresponds to the region where $\mathcal{G}<0$ (and thus $\mathcal{G}_R<0$), i.e.~the EoS is non-convex. The red area is the relativistic non-convex region (i.e.~$\mathcal{G}>0$ and $\mathcal{G}_R<0$). Non-causal regions are
   displayed in black.} 
\label{plot_PvsRho}
\end{figure}
%

%
\section{Numerical setup for non-convex BNS merger simulations} 
\label{sec:initial_d_and_method}
%
%
\subsection{Initial data}
\label{sub:initialdata}
We consider BNS configurations in quasi-equilibrium circular orbits that inspiral, merge and form  dynamical stable remnants lasting more than $\gtrsim 15\,\rm ms$. The binaries consist of two identical irrotational NS modeled by a piecewise polytropic representation of (several) nuclear EoS using seven pieces as in~\cite{Read:2008iy}. The initial data are computed using the {\tt LORENE} code~\cite{Gourgoulhon:2000nn,tg02,Lorene}.
The initial coordinate separation of the binary is $\sim 44.3\,\rm km$, and the rest-mass of each NS is $M_0=1.4M_\odot$.  The initial properties  of our binaries are summarized in Table~\ref{table:Iparamenters}. 
These representative EoSs broadly satisfy the current observational constraints on NS masses and radii~(see Fig.~\ref{plot_MvsR}).
For instance, all EoSs predict that the maximum (gravitational) mass configuration of an isolated spherical star is larger than $ 2M_\odot$, consistent with: i) $M_{\rm sph}^{\rm max}>2.072^{+0.067}_{-0.066}M_\odot$ from the NICER and XMM analysis of PSR J0740+6620~\cite{Riley:2021pdl}; ii) $M_{\rm sph}^{\rm max}>2.01^{+0.017}_{-0.017}M_\odot$ from the NANOGrav analysis of PSR J1614-2230~\cite{Fonseca:2016tux}; iii) $M_{\rm sph}^{\rm max}>2.01^{+0.14}_{-0.14}M_\odot$ from the pulsar timing analysis of PSR J0348+0432~\cite{Antoniadis:2013pzd}; and $M_{\rm sph}^{\rm max} >2.14^{+0.20}_{-0.18}M_\odot$ from the NANOGrav and the Green Bank Telescope~\cite{NANOGrav:2019jur}. However,
the LIGO/Virgo/KAGRA analysis of GW170817 predicts that the tidal deformability of a $1.4\,M_\odot$
NS is $\Lambda_{1.4}= 190^{+390}_{-120}$ at the  $90\%$ credible level~\cite{Abbott:2018exr}, and hence only a few EoSs in Table~\ref{table:Iparamenters} satisfy this constraint. Also note that the constraints imposed by LIGOScientific:2020zkf are more uncertain as there is not complete confirmation that the secondary of this compact binary coalescence is actually a NS~\cite{LIGOScientific:2020zkf}.  We stress that we consider these EoSs because, as shown in Fig.~\ref{plot_MvsR}, they span a large range of NS central rest-mass densities, radii and maximum gravitational masses for irrotational NSs, allowing us to probe the impact of a non-convex dynamics on the GW spectrum
of BNS mergers.

%
%
\subsection{Simulations}
\label{sub:methods}
Much of the numerical approach employed here has been extensively discussed in previous work (see e.g.~\cite{Werneck:2022exo,Etienne:2015_co}). Therefore, in the following we only summarize the basic aspects, referring the reader to those references for further details and code tests.
%
\begin{figure}
\includegraphics[width=1\linewidth]{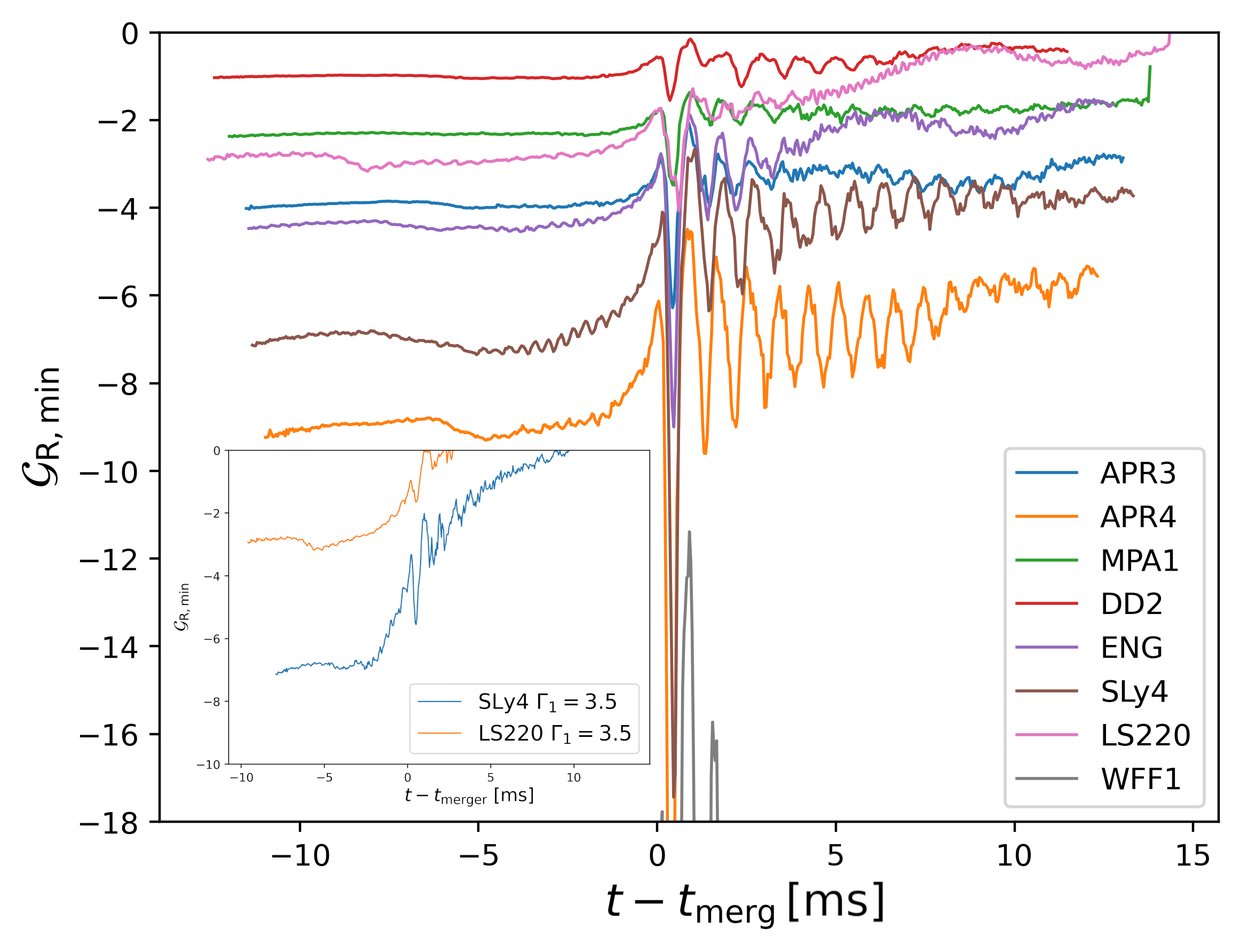}
  \caption{Minimum value of the relativistic fundamental derivative $\mathcal{G}_R$ vs coordinate time for some selected EoSs, evolved with the GGL EoS setting $\Gamma_1=3.0$ (see Table~\ref{table:Results}). The inset displays the evolution of $\mathcal{G}_R$ for two configurations evolved with $\Gamma_1=3.5$. Notice that the coordinate time has been shifted to the merger time.}
\label{plot:G_vs_time}
\end{figure}
%
%
\subsubsection{Evolution}
We carry out the simulations employing the well-tested, publicly available {\tt IllinoisGRMHD} code~\cite{Werneck:2022exo,Etienne:2015_co} embedded in the {\tt Einstein Toolkit} infrastructure~\citep{Loffler:ET}, employing the {\tt Carpet} code~\cite{Carpet} for the moving-box mesh capability. The code evolves the Baumgarte–Shapiro–Shibata–Nakamura (BSSN) equations for the spacetime fields~\citep{Baumgarte:1998_BS,Shibata:1995_SN} coupled with puncture gauge conditions cast in first-order form. We use fourth-order spatial stencils. We set the damping coefficient appearing in the shift condition to $1/M_{\rm ADM}$. Fifth-order Kreiss-Oliger dissipation~\cite{Kreiss89} is also added in the BSSN evolution equations. 
The {\tt IllinoisGRMHD} code evolves the matter fields using the equations of ideal, general-relativistic MHD, which are cast in a conservative scheme using the Valencia formulation~\citep{Anton:2006,Valencia_For,Font:2008}, via a high-resolution shock-capturing (HRSC) technique~\cite{Etienne:2015_co} that employs the piecewise parabolic method (PPM)~\cite{COLELLA1984174} coupled to the Harten, Lax, and van Leer (HLL) approximate Riemann solver~\cite{doi:10.1137/1025002}.  The time integration is performed  via the method of lines using a fourth-order accurate, Runge-Kutta integration scheme with a Courant-Friedrichs-Lewy factor of $0.45$.  

We notice that, if the approximate Riemann solver for the system of hydrodynamics equations closed with a non-convex EoS induces enough numerical viscosity to allow the formation of compound waves, then the resulting numerical scheme is stable (see~e.g.~\cite{1996ShWav...6..241A,GUARDONE200250,10.1063/1.4851415}). In particular, the use of the HLL solver allows us to resolve compound waves~\cite{Ibanez:2017xrx}.

%
%
\subsubsection{Grid structure}
In all simulations we use three sets of nested refinement boxes centered on each star and on the center of mass of the binary. Each of them contains six boxes that differ in size and in resolution by factors of two. When two boxes overlap they are replaced by a common box centered on the center of mass of the BNS. The finest box around the NS has a half-side length of $1.25\,R_{\rm NS}$, where $R_{\rm NS}$ is the radius of the NS. The finest level has a resolution of $\sim 220\,\rm m$ and resolves the star by $\sim 45$ grid point across its radius. We place the outer boundary at $\sim 1600\,\rm km$. 
%
%
\subsubsection{EoS for the dynamical evolution}
The cold EoS listed  in~Table~\ref{table:Iparamenters} are adequate to model the NS prior to merger. However, during merger considerable shock heating increases the internal energy and temperature to values over 10 MeV. Such high temperatures provide additional pressure
support that may alter the structure and evolution of the remnant. To account for this and following common practice, we adopt an EoS that has both a thermal and a cold contribution. The total pressure can be expressed as
\begin{equation}
p=p_{\rm th} + p_{\rm cold}\,,
\label{eq:pressure_thermal}
\end{equation}
where $p_{\rm cold}=\kappa_{i}\,\rho_0^{\Gamma_i}$, with $\kappa_i$ and $\Gamma_i$ the corresponding polytropic constant and the polytropic exponent in the rest-mass density range $\rho_{0,i-1}\leq \rho_0\leq \rho_{0,i}$, respectively (see~e.g.~\cite{Read:2008iy}), and  the thermal pressure is given by
\begin{equation}
p_{\rm th}=(\Gamma-1)\,\rho_0\,(\epsilon-\epsilon_{\rm cold})\,,
\label{eq:P-th}
\end{equation}
where
%
%
\begin{center}
  \begin{table}
    \caption{Summary of key results. First three columns display the times and frequencies at merger reported  in milliseconds, while the three last columns report the frequency in kiloHertz. Superindices denote the EoS used during the evolution: i)${\rm const\,\Gamma}$ denotes $\Gamma=1.8$; ii) ${\rm GGL,3.0}$  denotes GGL with $\Gamma_{\rm th}=1.8$ and $\Gamma_1=3.0$; and iii) ${\rm GGL,3.5}$  denotes GGL with $\Gamma_{\rm th}=1.8$ and $\Gamma_1=3.5$.  We define the merger time as the peak GW amplitude. A dash symbol denotes "not applicable".
      \label{table:Results}}
    \begin{tabular}{ccccccccc}
      \hline
      \hline
       EOS    & $t_{\rm mer}^{\rm const\,\Gamma}$   & $t_{\rm mer}^{\rm GGL,3.0}$ & $t_{\rm mer}^{\rm GGL,3.5}$  &  
       $f_{\rm peak}^{\rm {\rm const\,\Gamma}}$ 
       &$f_{\rm peak}^{\rm GGL,3.0}$ &$f_{\rm peak}^{\rm GGL,3.5}$  \\
      \hline
       WFF1   &   19.70   & 9.00   &-   & 3.56    & 3.44 &  -   \\
       WFF2   &   18.79   & 10.43  &-   & 3.35    & 3.15 &  -   \\
       APR4   &   18.52   & 10.96  &7.34& 3.33    & 3.05 &  2.40  \\
       SLy4   &   17.67   & 11.33  &7.91& 3.17    & 2.78 &  2.41   \\
       APR3   &   17.33   & 11.50  &-   & 2.99    & 2.77 &  -   \\
       ENG    &   17.14   & 11.43  &-   & 2.95    & 2.77 &  -   \\
       MPA1   &   16.31   & 11.97  &-   & 2.73    & 2.67 &  -   \\
       LS220  &   15.70   & 12.57  &9.60& 2.80    & 2.42 &  2.08  \\
       ALF2   &   15.73   & 11.76  &-   & 2.70    & 2.52 &  -   \\
       DD2    &   14.66   & 12.39  &9.67& 2.57    & 2.29 &  2.09   \\
       HShen  &   12.47   & 12.06  &-   & 2.20    & 2.14 &  -   \\
       GNH3   &   13.78   & 12.96  &-   & 2.48    & 2.29 &  -  \\
       MS1    &   11.69   & 11.52  &-   & 2.18    & 2.04 &  -   \\  
       \hline
    \end{tabular}
  \end{table}
\end{center}
\begin{equation}
\epsilon_{\rm cold} =-\int p_{\rm cold}\,d(1/\rho_0)\,.
\end{equation}
Following~\cite{Ibanez:2017xrx,Aloy_2019}, in order to incorporate non-convex thermodynamics in the evolution of the BNS initial data we employ the phenomenological GGL EoS. Therefore, the effective thermal index $\Gamma$ given by Eq.~(\ref{eq:Gamma_rho}) is used in the thermal part of the pressure, Eq.~(\ref{eq:P-th}). For comparison purposes with our previous simulations in~\cite{Guerra:2023} we set $\Gamma_{\rm th}=1.8$. The other three parameters of the GGL EoS are chosen such as the resulting EoS is causal, i.e. the speed of sound $c_{s(R)}< 1$. In particular, we  set $\rho_1=9.1\times 10^{14}\,\rm g\,cm^{-3}$, and $\Sigma=0.35\,\rho_1$. On the other hand, we consider different values of $\Gamma_1$ to study the impact of this parameter on the EoS and ultimately on the fate of the remnant. In our fiducial model we set $\Gamma_1=3.0$ and in some selected cases we set $\Gamma_1=3.5$.
%
\begin{figure}
    \includegraphics[width=1\linewidth]{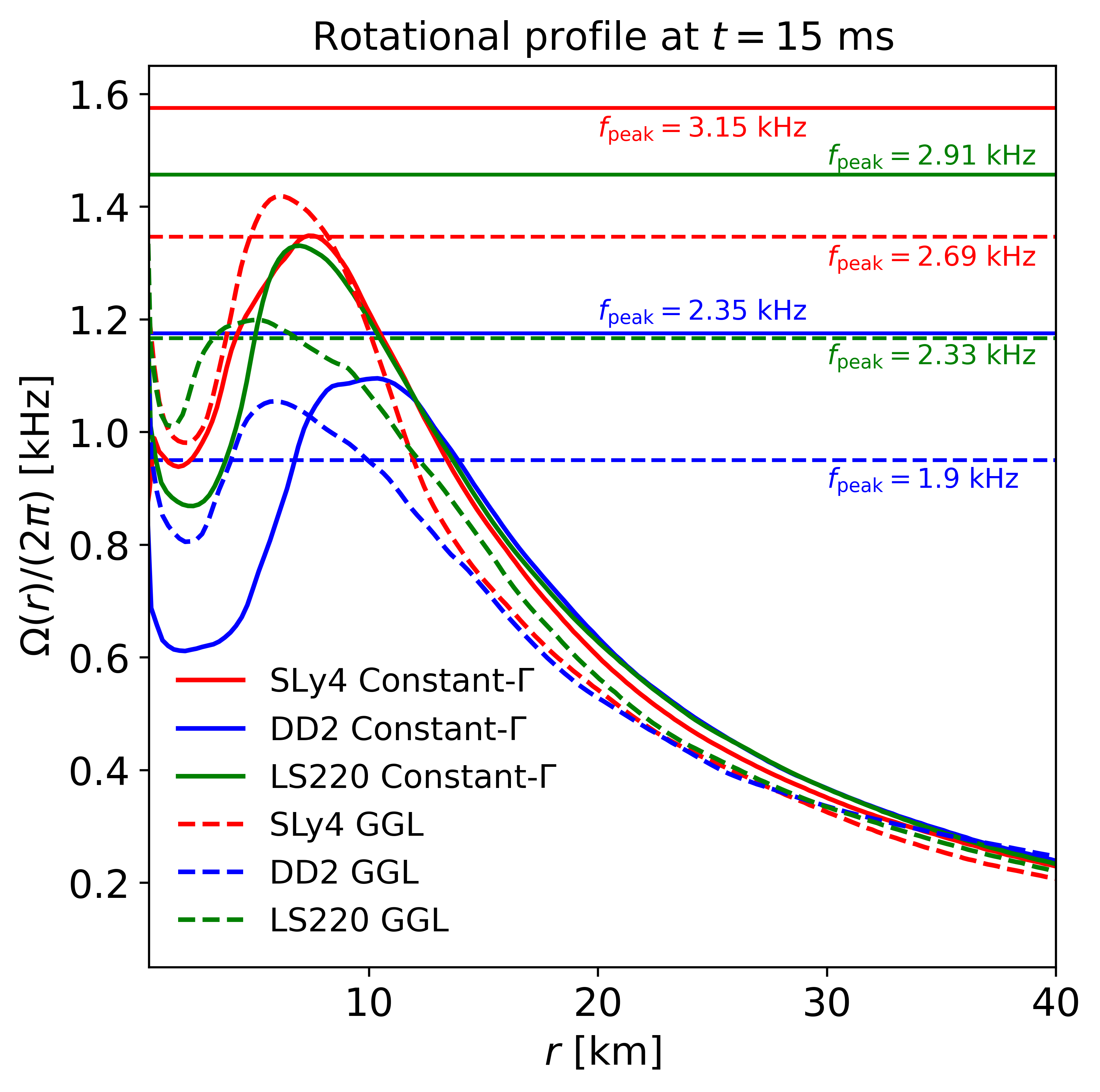}
  \caption{Rotational profiles of the binary remnant along the coordinate radius $r$  at $t\sim 15\,\rm ms$ after merger for the EoS indicated in the legend. Binaries with non-convex (convex) dynamics are displayed with dashed (continuous) lines. Horizontal lines show  the pattern speed frequencies of the main modes ($f_{\rm peak}/2$) for each model. The colors of the horizontal lines are the same as for the EoS labels. Similar profiles are observed for all other EoS in Table~\ref{table:Iparamenters}.} 
\label{plot:Omega}
\end{figure}

Fig.~\ref{plot_PvsRho} displays the convexity behavior of the GGL EoS on the $p-\rho_0$ plane. For typical values of the rest-mass density of a NS ($\rho_0\lesssim 10^{15}\,\rm g\,cm^{-3}$) the GGL EoS is convex (blue regions), i.e.~the fundamental derivatives~$\mathcal{G}$ and $\mathcal{G}_R$ are both positive. By contrast, in the green region~$\mathcal{G}\lesssim 0$ and so is the relativistic fundamental derivative. This is the non-convex region  of the EoS. There is also a non-convex relativistic region displayed in red where $\mathcal{G}_R< 0$, although $\mathcal{G}> 0$.  Regions where the speed of sound becomes superluminal are shown in black. Our BNS simulations are tuned in such a way that their evolution takes place in the red and/or green regions. 
%
%
\begin{figure*}
 \includegraphics[scale=01.05]{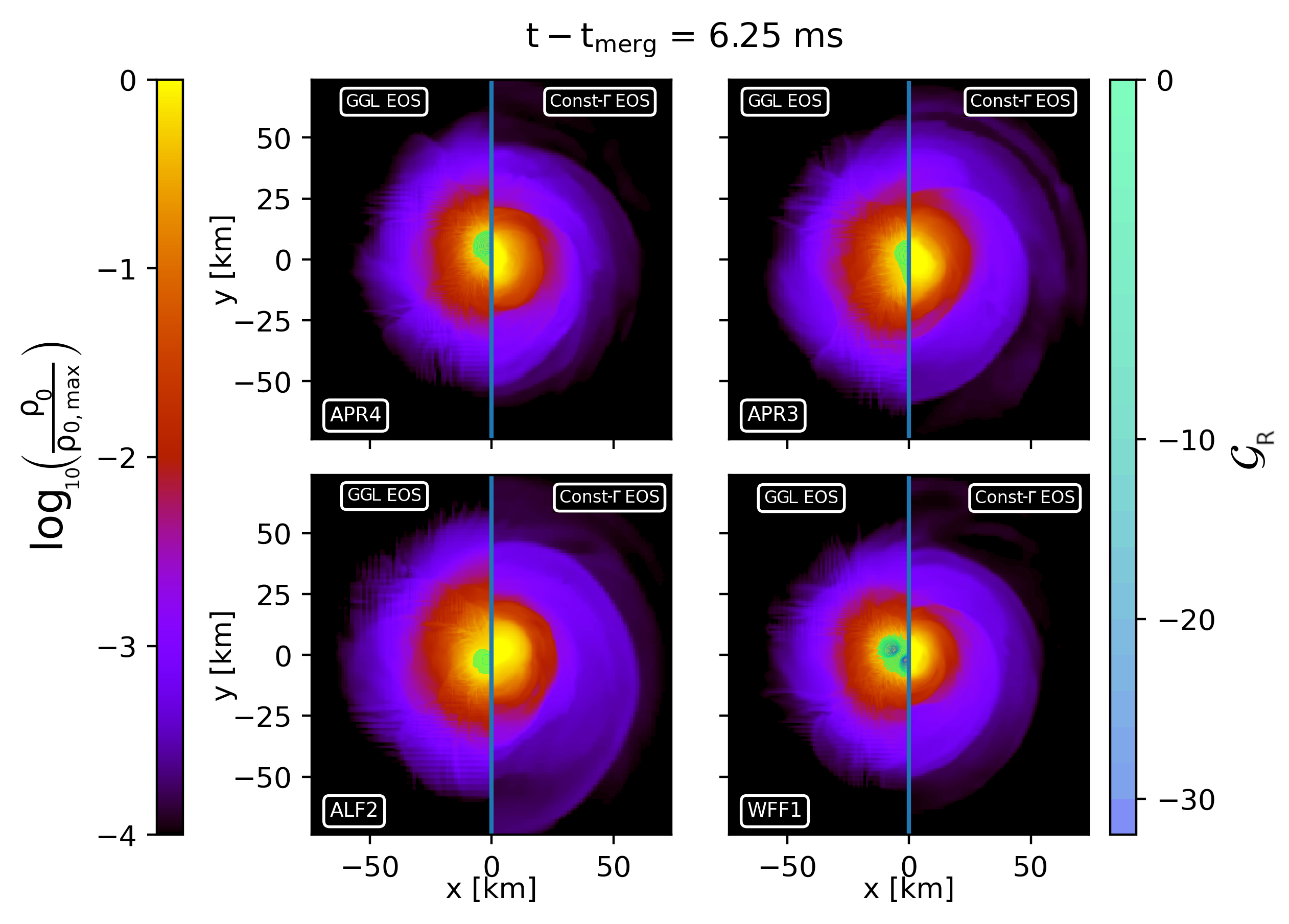}
    \caption{Rest-mass density $\rho_0$, normalized to its initial maximum value $\rho_{0,\rm max}\sim 10^{14.8}\,\rm g\,cm^{-3}$ (see Table~\ref{table:Iparamenters} for all cases), in log scale of the BNS remnant at $t\sim 6\,\rm ms$ following merger for some of the EoSs listed in Table~\ref{table:Results}. Negative values of the relativistic fundamental derivative are displayed in greenish (cold) color. Each panel shows a side-by-side comparison between the quasisteady configuration of the remnant evolved using either the phenomenological GGL EoS (left) or the constant $\Gamma-$law EoSs (right).} 
    \label{plot:rho_max}
\end{figure*}
%

%
\begin{figure*}
    \includegraphics[scale=1.15]{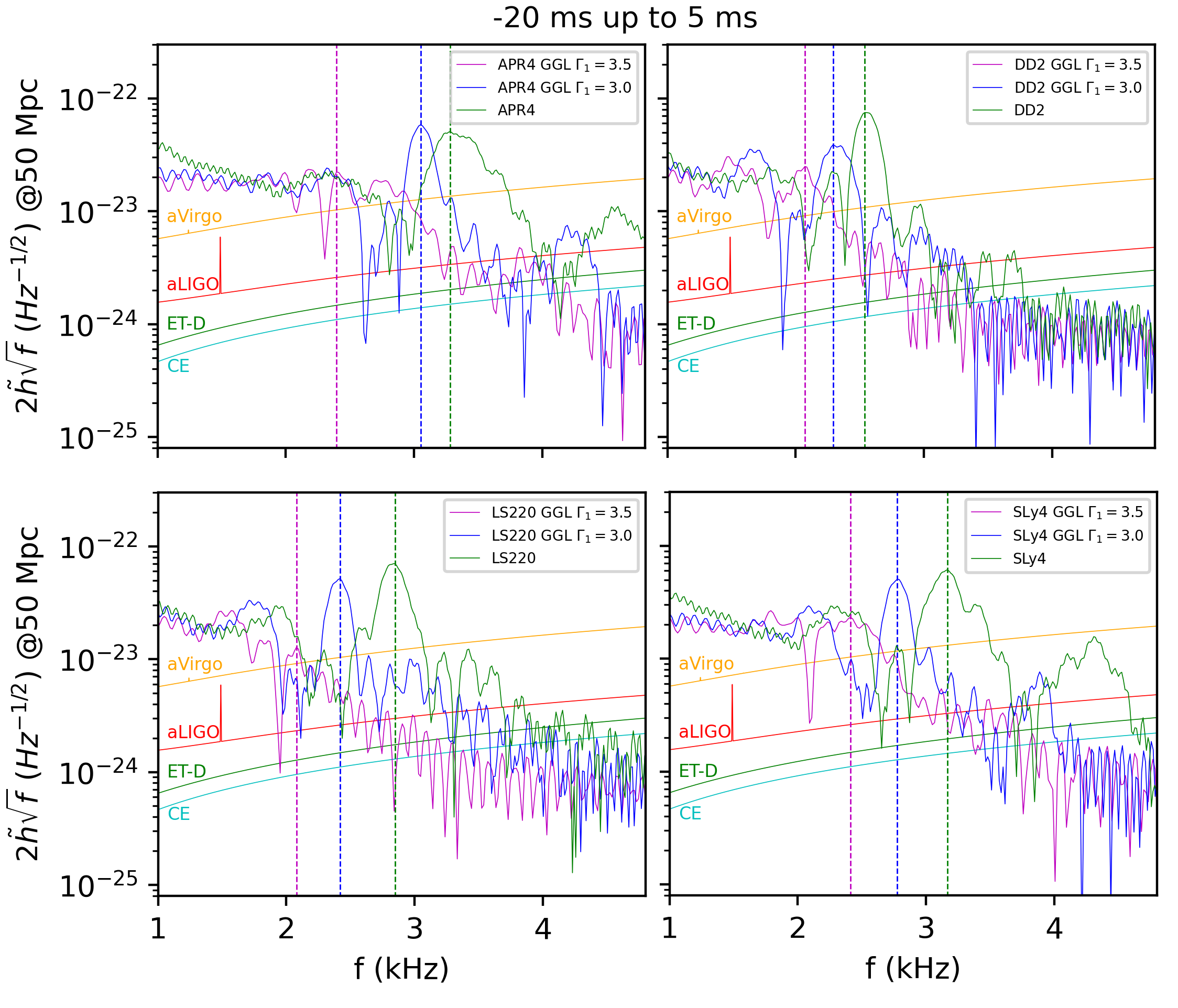}
  \caption{GW spectra of both GGL and constant$-\Gamma$ law binaries at $50$ Mpc with optimal orientation and with a time window $t-t_{\rm mer}=[-20,5]\,\rm ms$ which emphasizes the contribution of the dominant spectral mode $f_{\rm peak}$ following merger.  The two spectra for the GGL EoS correspond to $\Gamma_1 =3.0$ and $\Gamma_1=3.5$.   Sensitivity curves of Advanced Virgo (aVirgo)~\cite{VIRGO:2014yos}, Advanced LIGO (aLIGO)~\cite{LIGOsens}, Einstein Telescope~(ET-D)~\cite{2011CQGra..28i4013H}, and Cosmic Explorer (CE)~\cite{2017CQGra..34d4001A} are also displayed. Vertical lines mark the location of the peak frequency $f_{\rm peak}$.} 
\label{plot:GW_spectrum}
\end{figure*}
%
%
\begin{figure}
\includegraphics[scale=0.55]{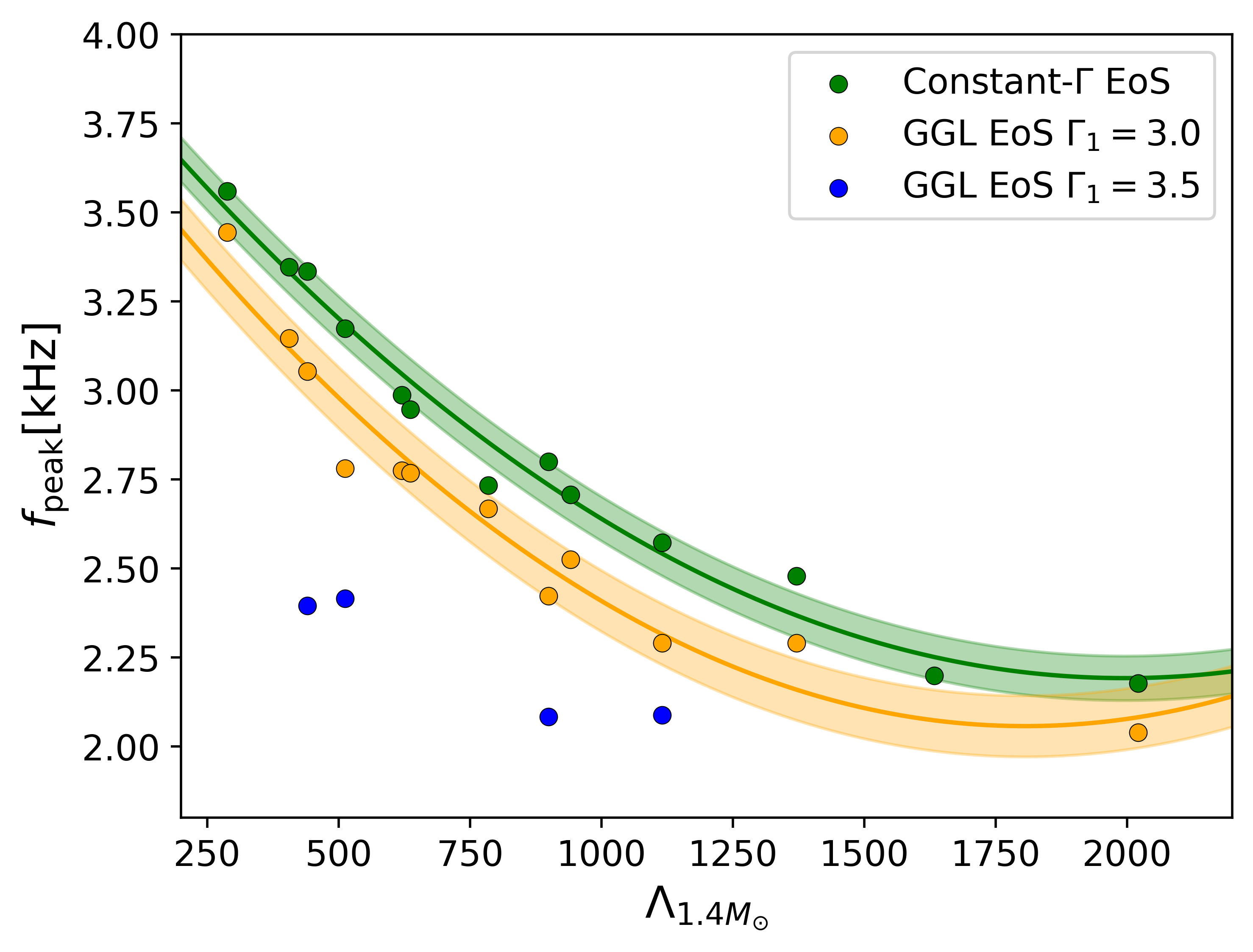}
    \caption{GW peak frequency of the post-merger remnant $f_{\rm peak}$ as a function of the tidal deformability $\Lambda$ for all simulations in Table~\ref{table:Results}. Each circle corresponds to one initial-data EoS, the corresponding colors indicating the EoS used in the evolutions (see legend). Solid curves display the quasi-universal relations given by Eqs.~(\ref{eq:peakGW}) and~(\ref{eq:peakGW_GGL}).  Colored regions represent the standard deviation of the corresponding fit.} 
    \label{plot:Final_plot}
\end{figure}

%

\section{Results}
\label{sec:discussion}

We turn now to describe the results of our simulations. We first note that for every non-convex model we consider (evolved with the GGL EoS), we also simulate the corresponding convex one, which is simply evolved using a constant value of the thermal index $\Gamma$ (simply achieved by setting $\Gamma_{\rm th}=\Gamma_1$ in Eq.~(\ref{eq:Gamma_rho})). The main quantitative results of this comparison, with regard to the time of merger and to the GW peak frequency at merger, are reported in Table~\ref{table:Results}.
%
%
\subsection{Morphology and dynamics}
The binaries start from a coordinate separation of $\sim 44.3\,\rm km$, which roughly corresponds to $\sim 6$ orbits before merger.  As GWs extract energy and angular momentum from the system, the coordinate orbital separation shrinks causing the stars to eventually merge forming a transient remnant with a lifetime $\gtrsim 15\,\rm ms$. 

During inspiral the dynamics for convex and non-convex EoS is similar although some differences appear. In particular, we observe that a non-convex dynamics tends to reduce the NS compactness compared to those with a convex dynamics. This behavior is somehow anticipated because during inspiral, the binary companion induces tidal forces that stretch the star out along the line connecting the centroids of the two stars. This effect triggers expansive shock waves in the bulk of a non-convex NS pushing out its outer layers, which in turn accelerates the merger. We observe that this behavior  depends on the EoS employed to build the initial data. The softer the EoS the shorter the merger time of the non-convex binary compared with that of binaries with convex dynamics. Binaries built with soft EoS, such as WFF1 or APR4, when evolved with the GGL EoS merge $\gtrsim 5\,\rm ms$ earlier than their counterparts evolved with a constant $\Gamma=1.8$ EoS. By contrast, binaries built with stiff EoS, such as DD2 or LS220, when evolved with the GGL EoS merge roughly at the same time ($\Delta t_{\rm mer}\lesssim 1\,\rm ms$) than their constant $\Gamma=1.8$ EoS counterparts (see Table~\ref{table:Results}).  These different dynamics can also be deduced from Fig.~\ref{plot:G_vs_time} which displays the
minimum value of the relativistic fundamental derivative $\mathcal{G}_R$ for binaries evolved with the GGL EoS and $\Gamma_1=3.0$. Soft EoS binaries tend to have smaller values of $\mathcal{G}_R$, and so are susceptible quite generally to developing strong expansive shock waves, that in turn puff the star out. By contrast, stiff EoS binaries tend to have larger values of $\mathcal{G}_R$, closer to zero, with expansive waves that only perturb the stars slightly. The dynamics of the latter mimics that of the convex binaries.

Following merger, a massive remnant forms with two dense cores rotating about each other. Depending on the compactness of the progenitors, and so on the value of the relativistic fundamental derivative~(see Fig.~\ref{plot:G_vs_time}), the two cores collide during the first $\lesssim 5\,\rm ms$ after merger giving birth to a  highly differentially rotating star wrapped in a Keplearian-like cloud of low density matter. These rotational profiles are displayed in Fig.~\ref{plot:Omega} at $t\sim 15$ ms after merger for some selected cases, but the behavior is similar for all models we have evolved. 
At this stage of the evolution only the convex version of EoS SLy4 and LS220 are still in corotation with the pattern speed frequency of the dominant quadrupolar mode (shown by the horizontal lines). The remnant settles down shortly after that, i.e.~the maximum value of the rest-mass density reaches a quasisteady state.  In particular, we observe that the WFF1-GGL binary remnant settles down at $t_{\rm merg}\sim 5.5\,\rm ms$, the WFF1 remnant with $\Gamma=1.8$ at $t_{\rm merg}\,\sim 4.2\,\rm ms$, while both the DD2-GGL and DD2 with $\Gamma=1.8$~remnants settle down at about $t_{\rm merg}\,\sim 1.0\,\rm ms$.  

Fig.~\ref{plot:rho_max} displays the rest-mass density $\rho_0$ on the equatorial plane at $t\sim 6$ ms for some binary  remnants in Table~\ref{table:Results} (see legend at the bottom-left corner in each panel), normalized to its initial maximum value $\rho_{0,\rm max}\sim 10^{14.8}\rm\, g\,cm^{-3}$ (see~Table~\ref{table:Iparamenters} for all cases). The left half of each panel in this figure shows the snapshot of the evolution with the GGL EoS while the right half corresponds to the $\Gamma$-law Eos with $\Gamma=1.8$. Following~\cite{Kastaun:2016yaf,Ruiz:2021qmm}, we define the bulk of the remnant as the region enclosed by the rest-mass isodensity contour $\gtrsim 10^{12.5}\,\rm g\,cm^{-3}$.  During the early post-merger evolution with the GGL EoS, the bulk of the remnant undergoes a transient period where non-convex relativistic regions~($\mathcal{G}_R<0$), identified by the green shaded areas in Fig.~\ref{plot:rho_max},  expand and contract continuously. We observe  that depending on the softness of the EoS these regions, as expected, may spread covering a large part of the bulk of the remnant or be confined around its central core. For all cases, once the remnant settles down we observe that the non-convex regions remain bounded within $\rho_0\gtrsim 10^{13.7}\,\rm g\,cm^{-3}$ (see Fig.~\ref{plot:rho_max}).  We note that, for all EoS listed in Table~\ref{table:Results}, the remnants resulting from non-convex evolutions are in general more extended than those of the convex ones,  which implies that the presence of a non-convex dynamics {tends to increase the pressure support in the remnant. Therefore, one may conclude that non-convex dynamics hardens the EoS of the remnant}. 
Consistent with this, the angular velocity profiles shown in Fig.~\ref{plot:Omega} for some of our models indicate that non-convex remnants tend to rotate more rapidly at the core than convex ones. This behaviour applies to our full set of models. Notice that this effect has also been observed in remnants undergoing quark-hadron phase transitions~\cite{Radice:2017,Most:2018eaw}.
%
%
\subsection{Gravitational wave spectra}
Fig.~\ref{plot:GW_spectrum} displays the GW spectra of the binaries for four representative initial-data EoS (APR4, DD2, LS220, and SLy4). The spectra are computed assuming a distance to the source of  $50\,\rm Mpc$, optimal orientation and sky localization, and using a fixed time window of $2\,\rm ms$, from $20\, \rm ms$ before merger to $5\,\rm ms$ past merger.  Following~\cite{DePietri:2019mti}, we do not apply any windowing functions to obtain a cleaner separation of the contributions in the spectra. As a result, the FFT may include small artifacts due to the finite size of the time intervals. 
Each panel in the figure contains three spectra, one obtained with a convex EoS and the other two with the non-convex GGL EoS. To illustrate the impact of the free parameters of the GGL EoS, the last two spectra correspond to evolutions with $ \Gamma_1= 3.0$ and $\Gamma_1= 3.5$. The frequencies of the dominant mode $f_{\rm peak}$ for each of these three evolutions are indicated with vertical dashed lines in each plot of the figure. The precise values are reported in the last three columns of Table~\ref{table:Results} for all EoS.

During inspiral, the existence of $\mathcal{G}_R<0$ regions slightly modifies the maximum GW frequency which reaches a maximum value below $2.2\,\rm kHz$ depending on the initial-data EoS. Following merger, the GW amplitude and frequency  strongly depend on both the softness of the initial-data EoS and of the existence of non-convex regions. Our simulations reveal that the frequency $f_{\rm peak}$ of the dominant oscillation mode of the remnant for binaries with non-convex dynamics  is always smaller than that of their convex dynamics counterparts (i.e.~the peak frequency in non-convex evolutions shifts to the left in Fig.~\ref{plot:GW_spectrum}).   In the most extreme cases analyzed, a frequency shift of $\Delta f_{\rm peak}\sim 300\,\rm Hz$ was found for the stiff DD2 EoS evolved with the GGL EoS with $\Gamma_1=3.0$ and a corresponding shift of  $\Delta f_{\rm peak}\sim 1\,\rm kHz$ was found for the soft APR4 EoS with $\Gamma_1=3.5$. All peak frequency values are reported in Table~\ref{table:Results} for all cases.

It is well known that the GW spectra of the post-merger remnant strongly depend on the EoS~(see~e.g.~\cite{Bauswein:2015yca,Vretinaris:2019spn,Takami:2014tva}) and, in particular, it has been shown that  the radius of the NS remnant depends on the peak prequency as $R_{\rm NS}\sim f_{\rm peak}^{-2/3}$~\cite{Bauswein:2011tp}. As the  tidal deformability $\Lambda$ is basically a function of the chirp mass $\mathcal{M}$ and  $R_{\rm NS}$~\cite{De:2018uhw,Bose:2017jvk}, it has been shown that future observations of the post-merger GW dominant mode could be used in combination with ``quasi-universal" relations between $\Lambda$ and $f_{\rm peak}$~(see e.g.~\cite{Rezzolla:2016nxn,Topolski:2023ojc}) to constraint the EoS of the NS. Notice that third-generation GW observatories may measure  $f_{\rm peak}$ within an accuracy $\lesssim 30\,\rm Hz$~\cite{Clark:2014wua,Yang:2017xlf,Torres-Rivas:2018svp}. Besides, it has also been  suggested~\cite{Bauswein:2018bma,Blacker:2020}  that significant deviations of $f_{\rm peak}$ from these quasi-universal relations with $\Lambda$ may indicate the presence of a first-order phase transition in the EoS leading to the formation of merger remnants with a quark-matter core. In particular~\cite{Bauswein:2018bma,Blacker:2020} performed numerical studies of BNS mergers in which the stars were modeled with a hybrid DD2F-SF EoS that exhibits a strong first-order phase transition to deconfine quarks within the standard Maxwell approach~\cite{compose,Typel:2009sy}. Those simulations showed that there is a shift of up to $\Delta f_{\rm peak}~\sim 700\,\rm Hz$ with respect to the value of the frequency attained in  BNS simulations  modeled with the purely hadronic EoS DD2F, an EoS equivalent to DD2F-SF but {\it without} phase transitions. 

The frequency shifts and ``outliers'' in the $f_{\rm peak}-\Lambda$ quasi-universal relation found by \cite{Bauswein:2018bma,Blacker:2020} may be attributed to the existence of non-convex dynamics in the post-merger remnant discussed here. Fig.~\ref{plot:Final_plot} displays  the dominant frequency as a function of the tidal deformability $\Lambda$ for all evolutions listed in Table~\ref{table:Results}. As expected, when the BNS merger is simulated using a convex EoS (green circles) $f_{\rm peak}$ is correlated with $\Lambda$ in a quasi-universal (EoS-insensitive) manner according to
 \begin{eqnarray}
f_{\rm peak}^{\rm const\,\Gamma}&=&\left[(4.52\pm 0.79)\times 10^{-7} \,\Lambda^2\right.
\nonumber \\
&-&\left.(1.80\pm 0.18)\times 10^{-3}\,\Lambda +(3.99\pm 0.08)\right]\,\rm kHz\,.\nonumber \\
\label{eq:peakGW}
 \end{eqnarray}
However, when the system is subject to non-convex dynamics the peak frequency is more than $1-\sigma$  away from the above quasi-universal relation (yellow and blue circles in Fig.~\ref{plot:Final_plot}). We notice that a $1-\sigma$ deviation in our previous fit corresponds to $\sim 64\,\rm Hz$. This significant shift in  frequency can solely be attributed in our study to the presence of a non-convex dynamics and not to the presence of a physical process such as a phase transition in the binary remnant, as our GGL EoS does not account for such an effect, and neither to spurious artifacts induced by the numerical access to a tabulated EoS.

{Notice that, in contrast with the results reported in~\cite{Bauswein:2018bma} where the effects of the first-order phase transition {\it increase} the peak frequency of the dominant mode, the appearance of a non-convex dynamics {\it decreases} this frequency. This discrepancy is simply due to the effective stiffening of the EoS remnant associated with our specific choice of the free parameters of our phenomenological GGL EoS, Eq.~(\ref{eq:Gamma_rho}). However, the {\it magnitude} of the frequency shifts is, in both cases, comparable. We expect that a survey of parameters of the GGL EoS, in particular accounting for a softening of the EoS remnant, may not only reproduce the magnitude of the shifts but also reconcile the direction of the shift of the peak frequency with the values reported by~\cite{Bauswein:2018bma}.}
 
Interestingly, we  find that the outliers to the fitting formula (\ref{eq:peakGW}) corresponding to binaries evolved with the GGL EoS setting $\Gamma_1=3.0$ satisfy their own quasi-universal relation according to
\begin{eqnarray}
f_{\rm peak}^{\rm GGL}&=&\left[(5.4\pm 1.1)\times 10^{-7} \,\Lambda^2\right. 
\nonumber 
\\
&-&\left. (1.95\pm 0.25)\times 10^{-3}\,\Lambda +  (3.82\pm 0.12)\right]\,\rm kHz\,,\nonumber \\
\label{eq:peakGW_GGL}
 \end{eqnarray}
with a deviation of $\Delta f_{\rm peak}\lesssim 380\,\rm Hz$ with respect to binaries subject to convex dynamics.  This finding suggests that the appearance of non-convex regions might induce  a displacement of the quasi-universal relation on the $f_{\rm peak}-\Lambda$ plane for non-convex EoS. 

The effects of a non-convex dynamics on the post-merger GWs can be enhanced by fine-tuning the free parameters of the thermal index in Eq.~(\ref{eq:Gamma_rho}). In particular, by setting $\rho_1=9.1\times 10^{14}\,\rm g\,cm^{-3}$,  $\Sigma=0.35\,\rho_1$, and $\Gamma_1=3.5$ we obtain the largest shift in  $f_{\rm peak}$, namely $\sim 1\,\rm kHz$ (see Table~\ref{table:Results} and the blue markers in Fig.~\ref{plot:Final_plot}).

To further corroborate that the frequency shifts observed in our non-convex evolutions are due to the ``anomalous'' dynamics and not to the specific parameters of the GGL EoS of our fiducial model, we also simulate BNS mergers for the initial-data EoS DD2 with values of the GGL EoS parameters so that the evolution remains {\it convex} throughout (see~Appendix~\ref{sec:appendix} for details). In all cases we find that 
$10\,\rm Hz\lesssim\Delta f_{\rm peak}\lesssim 50\,\rm Hz$, i.e.~less than $1-\sigma$ away from the quasi-universal relation of Eq.~(\ref{eq:peakGW}) found for convex EoS. This suggests that the significant shifts in the peak frequency reported in this paper should be induced by the non-convex dynamics. 
  
%

\section{Discussion}
\label{sec:Conclusion}

Understanding the properties of matter beyond nuclear saturation density is essential for explaining GW observations of BNS mergers. Knowledge has been collected through combined experimental and theoretical efforts including EM and GW observations, nuclear physics experiments, and numerical simulations. The latter are distinctly driven by the ongoing GW observations of compact binary coalescences reported by the LIGO-Virgo-KAGRA (LVK) Collaboration~\cite{KAGRA:2021vkt} and by the expected significant increase in the rate of detections once third-generation detectors come online. Moreover, BNS mergers are the prime sources for multimessenger astronomy. EM, GW, and neutrino observations of these systems, in combination with theoretical and numerical work, will help advance our understanding of the origin of short gamma-ray bursts, r-process nucleosynthesis, or the nature of matter beyond nuclear saturation density. 

In the past few years numerical simulations of BNS mergers have advanced in the treatment of the thermodynamics of the system, encoded in the EoS of high-density matter. Current simulations incorporate microphysical, finite-temperature EoS tables constructed using ``tabulated'' data from observations and nuclear physics experiments~\citep{Oechslin:2006uk,Bauswein:2010dn,Sekiguchi:2011zd,Bauswein:2018bma,Most:2018eaw,Espino:2022mtb,Werneck:2022exo,Guerra:2023}. Some of the EoS employed also allow for phase transitions from nuclear hadronic matter into quark-gluon plasma or into matter phases containing exotic particles, processes that may modify the dynamics of the merger remnant and the GW emission. Numerical studies have searched for a first-order hadron-quark phase transition on the GWs~(see e.g.~\cite{Radice:2017,Bauswein:2018bma,Most:2018eaw,Blacker:2020,Liebling:2020dhf,Ujevic:2022nkr}). In particular~\cite{Bauswein:2018bma,Blacker:2020} reported a deviation of the quasi-universal relation between the tidal deformability $\Lambda$ and the peak GW frequency at merger $f_{\rm peak}$ if the EoS allows for a strong first-order phase transition to deconfined quarks. The presence of a phase transition may also affect the monotonic increase of the speed of sound with density, turning a convex dynamics into a non-convex one.

In this paper we have performed numerical-relativity simulations of BNS mergers subject to non-convex dynamics, allowing for the appearance of expansive shock waves and compressive rarefactions.
To this aim we have used a phenomenological non-convex EoS proposed in~\cite{Ibanez:2017xrx} and also used in~\cite{Aloy_2019}. The latter work showed that the appearance of non-convex dynamics during the gravitational collapse of uniformly rotating NS leaves a distinctive imprint on the GW signal, and served as a motivation for this study. Further motivation was gathered by our attempt to provide an explanation to the loss of the $\Lambda-f_{\rm peak}$ quasi-universal relation found by~\cite{Bauswein:2018bma,Blacker:2020} for EoS admitting a strong first-order phase transition. We have surveyed a number of BNS initial configurations modeled with a piecewise-polytropic representation of different (cold) nuclear EoS. Those have been subsequently evolved with a $\Gamma-$law EoS to allow for shock heating, considering two different possibilities for the adiabatic index $\Gamma$, either a constant value, which induces a convex (regular) dynamics, or  a variable index depending on the rest-mass density, which induces a non-convex (anomalous) dynamics.    

By comparing the two types of dynamics -- convex vs non-convex -- we have identified observable differences in the GW spectra of the remnant. In particular, we have found that non-convexity induces a significant shift  in the $\Lambda-f_{\rm peak}$ quasi-universal relation, of order $\Delta f_{\rm peak}\gtrsim 380\,\rm Hz$, with respect to that of binaries with convex dynamics. These values are similar {in magnitude} to those reported by~\cite{Bauswein:2018bma,Blacker:2020}, attributed however to a first-order phase transition from nuclear/hadronic matter to deconfined quark matter. We argue that the ultimate origin of the frequency shift is to be found in the presence of anomalous, non-convex dynamics in the binary remnant. 


The BNS merger simulations of~\cite{Bauswein:2018bma} comprise an extensive number of EoS. Their fiducial model is based on the temperature-dependent,
microscopic, hadron-quark hybrid EoS DD2F-SF of~\cite{Fischer:2018} and its nucleonic counterpart DD2F (with no phase transition). They consider different choices of parameters for the description of the quark phase, resulting in seven hybrid DD2F-SF EoS, which cover models with different onset densities and density jumps. In addition they also consider a representative sample of 15 EoS  describing purely hadronic models, three of which include a second-order phase transition to hyperonic matter, as well as the EoS ALF2 and ALF4 from~\cite{Read:2008iy}, which resemble models with a continuous transition to quark matter without a density jump. Out of all these EoS, the only ones departing from the tight quasi-universal scaling between $f_{\rm peak}$ and the tidal deformability are the seven hybrid DD2F-SF EoS. For this subset of hybrid EoS~\cite{Bauswein:2018bma} observed that the larger the density jump the more prominent the departure from the quasi-universal relation. However, the three EoS from their sample including a second-order phase transition to hyperonic matter were found to follow closely the $f_{\rm peak}-\Lambda$ relation similarly to purely nucleonic EoS (in agreement with early results from~\cite{Radice:2017}). The same scaling is found for EoS ALF2 and ALF4 in which the phase transition is continuous. In summary \cite{Bauswein:2018bma} conclude that only a sufficiently strong first-order phase transition can alter the postmerger GW signal in such a way that a measurable deviation from the $f_{\rm peak}-\Lambda$ relation occurs.

The fact that the scaling is lost only when the density jump is large enough might be, perhaps, an indication of possible numerical inaccuracies. In particular, the evaluation (through discretization) of high-order derivatives in tabulated dense-matter EoS across coexistence boundaries in phase transitions may introduce small-scale oscillations of numerical origin. This is visible in Fig.~\ref{plot:phase_tran} (showing the results for the DD2F-SF EoS employed by~\cite{Bauswein:2018bma,Blacker:2020}) and in Figs.~2 and 3 of~\cite{Aloy_2019}. We recall that those oscillations in the fundamental derivative are a numerical artifact due to the computation of the derivatives in ${\cal G}$ using a discrete EoS table. (See also~\cite{Vaidya:2015} for an additional example of the {\it numerical} loss of convexity associated with insufficient thermodynamic discretization in tabulated EoS.)  The effect of these oscillations is to spread spuriously the non-convex behaviour to points beyond the phase transition boundaries. In addition, as showed in~\cite{Aloy_2019}, many of the microphysical EoS investigated in that work display a sensitive reduction of the relativistic fundamental derivative as the baryon number density grows above 1 fm$^{-3}$. In that regime, even spurious small-scale oscillations may drive the relativistic fundamental derivative towards negative values. This would artificially trigger non-convex thermodynamics {\it of a numerical origin} in those regions with number densities above 1 fm$^{-3}$. As suggested by~\cite{Aloy_2019}, this artificial behavior could be attenuated using a finer number of data points in those regions of the EoS tables where the fundamental derivative displays large variations (i.e.~near the boundaries of first-order phase transitions), specially if these variations drive ${\cal G}$ to negative values.


The phenomenological GGL EoS employed in the simulations reported in this work can only be regarded as a toy model. Nevertheless, it has served the purpose of highlighting the potential relevance the development of non-convex dynamics may have on important observables in BNS mergers such as the GW emission (as was also shown by~\cite{Aloy_2019} in the context of gravitational collapse). 
{Notice that the free parameters of our EoS have been chosen such that the dynamics of the binary is always non-convex. In particular, for all our simulations $\Gamma_1> \Gamma_{\rm th}$ which induces a stiffer EoS compared to the constant $\Gamma-$law EoS (convex models).  One might also consider a softer GGL EoS by choosing $\Gamma_1< \Gamma_{\rm th}$ and analyze the impact of the EoS stiffness on the non-convex dynamics. Such a study has not been attempted in this work.
We hypothesize that our results are robust against changes in the stiffness of the phenomenological GGL EoS as long as the resulting dynamics is non-convex, since the latest is responsible for hardening the EoS of the binary remnant.} We stress that our GGL EoS does not account for phase transitions and is not affected by potential spurious artifacts induced by the numerical access to a tabulated EoS as all derivatives can be computed analytically. A natural extension of this work will be to revisit these simulations using actual microphysical EoS allowing for such non-convex dynamics. Those could also be carried out in combination with the analytic model for modelling phase transitions in tabulated EoS recently reported by~\cite{Berbel:2023}.

%
\acknowledgments
We thank Jos\'e Mar\'{\i}a Ib\'a\~nez for a careful reading of the manuscript and for providing insightful comments. Work supported by the Generalitat Valenciana  (grants CIDEGENT/2021/046 and Prometeo CIPROM/2022/49), and by the Spanish Agencia Estatal de  Investigaci\'on (grants PID2021-125485NB-C21 funded by MCIN/AEI/10.13039/501100011033,  PRE2019-087617, and ERDF A way of making Europe). Further support has been provided by the EU's Horizon 2020 Research and Innovation (RISE) programme H2020-MSCA-RISE-2017 (FunFiCO-777740) and  by  the  EU  Staff  Exchange  (SE)  programme HORIZON-MSCA-2021-SE-01 (NewFunFiCO-101086251). We acknowledge computational resources and technical support of the Spanish Supercomputing Network through the use of MareNostrum at the Barcelona Supercomputing Center (AECT-2023-1-0006). This work has used the following open-source packages: \textsc{NumPy}~\cite{harris:2020}, \textsc{SciPy}~\cite{scipy:2020}, \textsc{Scikit-learn}~\cite{scikit-learn}, and \textsc{Matplotlib}~\cite{Hunter:2007}. Finally,  we thank the authors of~\cite{DePietri:2019mti}
for making  available the \textsc{PyCactus} python post-process package for analyzing output of \textsc{EisteinToolkit}.
%
%
\appendix
\section{Impact of $\Gamma$ on $f_{\rm peak}$ }
\label{sec:appendix}
Thermal effects to account for shock heating during merger are modelled in our simulations using a hybrid approach, in which the EoS has a cold part and a thermal part (see Eq.~(\ref{eq:pressure_thermal})). The value of the thermal index $\Gamma$ appearing in the ideal-gas-like part of the pressure can be chosen somewhat arbitrarily, as long as it lays in the range $1\leqslant \Gamma\leqslant 2$ due to
causality constraints~(see~e.g.~\cite{Takami:2014tva}). However, the choice of $\Gamma$
controls the thermal pressure produced during and after merger (see Eq.~(\ref{eq:P-th})). In this Appendix we address if the observed shift on $f_{\rm peak}$ could be triggered by changes in the value of $\Gamma$. To do so we evolve the BNS built with the initial-data DD2 EoS using both the constant $\Gamma-$law and the GGL EoS using the parameters in Table~\ref{table:checks}.  We tune these parameters such that the dynamics of the system remains convex during the whole evolution. In all cases we find that the shift in the frequency of the dominant mode is $\Delta f_{\rm peak}\lesssim 50\,\rm Hz$ (see last column in Table~\ref{table:checks}). These results imply that the changes on the GW spectra discussed in this work are triggered by the  non-convex dynamics of the BNS mergers.
%
\begin{table}[t]
\caption{Summary of the parameters of the GGL EoS used to evolve a BNS built with the initial-data DD2 EoS. The choice of parameters guarantees a convex dynamics in all cases. The last column reports the shift on the peak  frequency with respect to our fiducial $\Gamma_{\rm th}=1.8$, $\Gamma_1=3.0$ case. Here $\rho_0$ denotes the initial value of the central density $\rho_0=5.70\times 10^{14}\,\rm g\,cm^{-3}$. A dash symbol
denotes ``not applicable''.}
\begin{tabular}{ccccc} \hline 
$\Gamma_{\rm th}$ & $\Gamma_{1}$ & $\Sigma$       & ${\rho_{1}}/{\rho_{0}}$ & $\Delta f_{\rm peak} [\rm Hz]$ \\ \hline 
1.36          & 1.36            & -              & -                           & 13                  \\
1.8           & 1.36         & 0.3$\rho_{1}$ & 0.69                         & 26                 \\
1.8           & 2.0          & 0.35$\rho_{1}$ & 0.91                        & 53                   \\
2.0           & 2.0            & -              & -                           & 40                    \\
\hline
 \\ \end{tabular}
\label{table:checks}
\end{table}

\bibliographystyle{apsrev4-1}
\bibliography{references}
\end{document}

%% file: kigou.tex
\newcommand{\beq}{\begin{equation}} 
\newcommand{\eeq}{\end{equation}} 
\newcommand{\beqn}{\begin{eqnarray}} 
\newcommand{\eeqn}{\end{eqnarray}}

\newcommand{\zD}{{\raise1.0ex\hbox{${}^{\ \circ}$}}\!\!\!\!\!D}
\newcommand{\alone}{{\raise0.5ex\hbox{${}^{\ 1}$}}\!\!\!\!\alpha}

\newcommand{\nalam}{\mathrel{\raise0.9ex\hbox{$^\lambda$}\mkern-14mu
\lower0.0ex\hbox{$\nabla$}}}

\newcommand{\zeroD}{{\raise1.0ex\hbox{${}^{\ \circ}$}}\!\!\!\!\!D}

\newcommand{\zLap}{{\raise1.0ex\hbox{${}^{\ \circ}$}}\!\!\!\!\Delta}
\newcommand{\zna}{{\raise1.0ex\hbox{${}^{\ \circ}$}}\!\!\!\!\!\nabla}
\newcommand{\zS}{{\raise1.0ex\hbox{${}^{\ \circ}$}}\!\!\!\!\!S}

